# An Adaptive parameter free data mining approach for healthcare application

Prof. Dipti Patil[#1]
# Asst. Professor, Computer Engg. Dept, MITCOE Pune, India

Bhagyashree Agrawal[#2]
#Computer Engg. Dept, MITCOE Pune, India

Snehal Andhalkar[*3]
#Computer Engg. Dept, MITCOE Pune, India

Richa Biyani[#4]
#Computer Engg. Dept, MITCOE Pune, India

Mayuri Gund[#5]
#Computer Engg. Dept, MITCOE Pune, India

Dr. V.M.Wadhai[#6]
#Professor, Computer Engg. Dept, MITCOE Pune, India

*Abstract*—**In today's world, healthcare is the most important factor affecting human life. Due to heavy work load it is not possible for personal healthcare. The proposed system acts as a preventive measure for determining whether a person is fit or unfit based on his/her historical and real time data by applying clustering algorithms viz. K-means and D-stream. Both clustering algorithms are applied on patient's biomedical historical database. To check the correctness of both the algorithms, we apply them on patient's current biomedical data. The Density-based clustering algorithm i.e. the D-stream algorithm overcomes drawbacks of K-means algorithm. By calculating their performance measures we finally find out effectiveness and efficiency of both the algorithms.**

## I. INTRODUCTION

Today the health care industry is one of the largest industries throughout the world. It includes thousands of hospitals, clinics and other types of facilities which provide primary, secondary & tertiary levels of care. The delivery of health care services is the most visible part of any health care system, both to users and the general public [2]. A health care provider is an institution or person that provides preventive, curative, promotional or rehabilitative health care services in a systematic way to individuals, families or community. The physiological signals such as SpO2, ABPsys, ABPdias, HR affects person's health. In health care the data mining is more popular and essential for all the healthcare applications. In healthcare industry having the more amounts of data, but this data have not been used properly for the application. In this health care data is converted in to the useful purpose by using the data mining techniques [1].

The data mining is the process of extracting or mining the knowledge from the large amounts of data, database or any other data base repositories. The main purpose of the data mining is to find the hidden knowledge from the data base. In health care industry, the data having some unwanted data, missing values and noisy data. This unwanted data will be removed by using preprocessing techniques in data mining. Preprocessing is the process of removing noise, redundant data and irrelevant data. After the preprocessing the data will be used for some useful purpose. In recent years different approaches are proposed to overcome the challenges of storing and processing of fast and continuous streams of data.

Data stream can be conceived as a continuous and changing sequence of data that continuously arrive at a system to store or process. Traditional OLAP and data mining methods typically require multiple scans of the data and are therefore infeasible for stream data applications. Whereby data streams can be produced in many fields, it is crucial to modify mining techniques to fit data streams. Data stream mining has many applications and is a hot research area[3]. Data stream mining is the extraction of structures of knowledge that are represented in the case of models and patterns of infinite streams of information. These data stream mining can be used to form the clusters of medical health data. This paper proposed two main clustering algorithms namely, K-means algorithm and density based clustering.

The K-means clustering algorithm is incompetent to find clusters of arbitrary shapes and cannot handle outliers. Further, they require the knowledge of k and user-specified time window. To address these issues, D Stream, a framework for clustering stream data using a density-based approach. The algorithm uses an online component which maps each input data record into a grid and an offline component which computes the grid density and clusters the grids based on the density. The algorithm adopts a density decaying technique to capture the dynamic changes of a data stream. Exploiting the intricate relationships between the decay factor, data density and cluster structure, our algorithm can efficiently and effectively generate and adjust the clusters in real time. Further, a theoretically sound technique is developed to detect and remove sporadic grids mapped to by outliers in order to dramatically improve the space and time efficiency of the system. The technique makes high-speed data stream clustering feasible without degrading the clustering quality. The experimental results show that our algorithm has superior quality and efficiency, can find clusters of arbitrary shapes, and can accurately recognize the evolving behaviors of real-time data streams [4].

## II. RELATED WORK

All Several health care projects are in full swing in different universities and institutions, with the objective of providing more and more assistance to the elderly. The application of data clustering technique for fast retrieval of relevant information from the medical databases lends itself into many different perspectives.

Health Gear: a real-time wearable system for monitoring and analyzing physiological signals [5] is a real-time wearable system for monitoring, visualizing and analyzing physiological signals. This system focused on an implementation of Health Gear using a blood oximeter to monitor the user's blood oxygen level and pulse while sleeping. The system also describes two different algorithms for automatically detecting sleep apnea events, and illustrates the performance of the overall system in a sleep study with 20participants. A Guided clustering Technique for Knowledge Discovery – A Case Study of Liver Disorder Dataset, [6] presents an experiment based on clustering data mining technique to discover hidden patterns in the dataset of liver disorder patients. The system uses the SOM network's internal parameters and k-means algorithm for finding out patterns in the dataset. The research has shown that meaningful results can be discovered from clustering techniques by letting a domain expert specify the input constraints to the algorithm.

Intelligent Mobile Health Monitoring System (IMHMS), [7] Author proposed the system which can provide medical feedback to the patients through mobile devices based on the biomedical and environmental data collected by deployed sensors. The system uses the Wearable Wireless Body/Personal Area Network for collecting data from patients, mining the data, intelligently predicts patient's health status and provides feedback to patients through their mobile devices.

The patients will participate in the health care process by their mobile devices and thus can access their health information from anywhere any time. But actual implementation of data mining framework for decision support system is not done.

Real-Time analysis of physiological data to support medical applications [8], proposed a flexible framework to perform real-time analysis of physiological data and to evaluate people's health conditions. Patient or disease-specific models are built by means of data mining techniques. Models are exploited to perform real time classification of physiological signals and continuously assess a person's health conditions. The proposed framework allows both instantaneous evaluation and stream analysis over a sliding time window for physiological data. But dynamic behavior of the physiological signals is not analyzed also the framework is not suitable for ECG type of signals.

Performance of Clustering Algorithms in Healthcare Database [1], proposed a framework where they used the heart attack prediction data for finding the performance of clustering algorithm. In final result shows the performance of classifier algorithm using prediction accuracy and the visualization of cluster assignments shows the relation between the error and the attributes. The comparison result shows that, the make density based clusters having the highest prediction Accuracy.

## III. METHOD

We present a framework that will perform clustering of dataset available from medical database effective manner. The flow of the system is depicted in Fig.1.

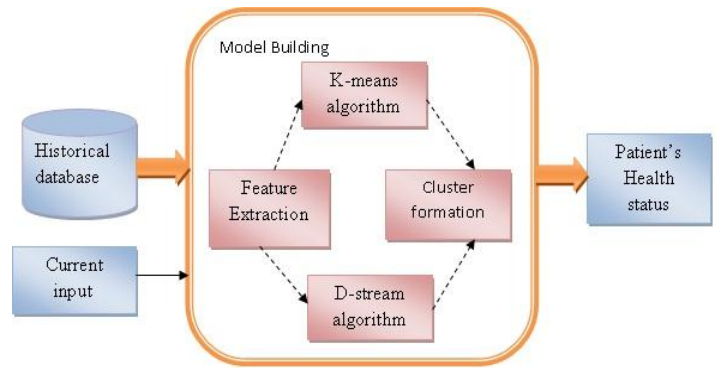

Figure 1.Flow of the system

The target is to cluster the patient's records into different groups with respect to the test report attributes which may help the clinicians to diagnose the patient's disease in efficient and The evaluation steps are the following-

*1) Data set collection*

The data set contains 7 attributes, SpO2, ABPsys, ABPdias, HR, heredity, obesity, cigarette smoking. These attributes are the risk factors that can help in predicting the patient's health status. Attributes such as SpO2, ABPsys, ABPdias, HR can be collected form MIMIC database [9] and the other attributes are influenced by the person's behavior. These all attributes values are discrete in nature .The dataset will be in preprocessed format.

*2) Model Building*

In model building phase features of the available data will be extracted and then clustering algorithm will be applied on extracted features.

*A. Feature Extraction*

For each physiological signal x among the X monitored vital signs, we extract the following features [8].

*1) Offset*

The offset feature measures the difference between the current value x(t) and the moving average (i.e., mean value over the time window). It aims at evaluating the difference between the current value and the average conditions in the recent past.

*2) Slope*

The slope function evaluates the rate of the signal change. Hence, it assesses short-term trends, where abrupt variations may affect the patient's health.

*3) Dist*

The dist feature measures the drift of the current signal measurement from a given normality range. It is zero when the measurement is inside the normality range.

### B. Risk Components

The signal features contribute to the computation of the following risk components.

*1) Sharp changes*

The $z_1$ component aims at measuring the health risk deriving from sharp changes in the signal (e.g., quick changes in the blood pressure may cause fainting)

*2) Long-term trends*

The $z_2$ component measures the risk deriving from the h weighted offset over the time window. While $z_1$ focuses on quick changes, $z_2$ evaluates long-term trends, as it is offset-based.

*3) Distance from normal behavior*

The $z_3$ component assesses the risk level given by the distance of the signal from the normality range. A patient with an instantaneous measurement outside the range may not be critical, but her/his persistence in such conditions contributes to the risk level

From above risk components, risk functions and global risk components will be calculated. These values will be further used in clustering algorithms as an input for cluster formation.

### C. Cluster formation

The proposed flow of the system uses two algorithms K-means and D-stream. The comparison between two clustering algorithms will be performed using the above described attributes.

#### K-MEANS ALGORITHM

1) The algorithm is composed of the following steps: [10] Place K points into the space represented by the objects that are being clustered. These points represent initial group centroids. Assign each object to the group that has the closest centroid. When all objects have been assigned, recalculate the positions of the K centroids.
Repeat Steps 2 and 3 until the centroids no longer move. This produces a separation of the objects into groups from which the metric to be minimized can be calculated.

Figure 2. Algorithm K-means algorithm

#### D-STREAM ALGORITHM

The D-stream algorithm is explained as follows [4]
1. procedure D-Stream
2. Tc = 0;
3. Initialize an empty hash table grid list;
4. while data stream is active do
5. read record x = (x1, x2, · · · , xd);
6. determine the density grid g that contains x;
7. if(g not in grid list) insert g to grid list;
8. update the characteristic vector of g;
9. if tc == gap then
10. call initial clustering(grid list);
11. end if
12. if tc mod gap == 0 then
13. detect and remove sporadic grids from grid list;
14. call adjust clustering(grid list);
15. end if
16. tc = tc + 1;
17. end while
18. end procedure

Figure 3: The overall process of D-Stream.

The overall architecture of D-Stream, which assumes a discrete time step model, where the time stamp is labeled by integers 0, 1, 2, · · · , n, · · · . D-Stream has an online component and an offline component. The overall algorithm is outlined in Figure 1.

For a data stream, at each time step, the online component of D-Stream continuously reads a new data record, place the multi-dimensional data into a corresponding discretized density grid in the multi-dimensional space, and update the characteristic vector of the density grid (Lines 5-8 of Figure 3). The density grid and characteristic vector are to be described in detail later. The offline component dynamically adjusts the clusters every gap time steps, where gap is an integer parameter. After the first gap, the algorithm generates the initial cluster (Lines 9-11). Then, the algorithm periodically removes sporadic grids and regulates the clusters (Lines 12-15).

D-Stream partitions the multi-dimensional data space into many density grids and forms clusters of these grids. This concept is schematically illustrated in Figure 4.

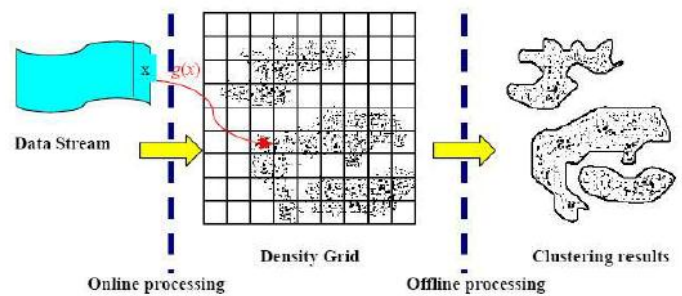

Figure 4. Illustration of the use of density grid.

The input data has d dimensions, and each input data record is defined within the space

$$S = S_1 \times S_2 \times \cdots \times S_d, \quad \ldots\ldots (1)$$

where Si is the definition space for the ith dimension.

In D-Stream, we partition the d−dimensional space S into density grids. Suppose for each dimension, its space Si, I = 1, · · · , d is divided into pi partitions as

$$S_i = S_{i,1} \bigcup S_{i,2} \bigcup \cdots \bigcup S_{i,p_i}, \quad \text{-----(2)}$$

then the data space S is partitioned into N density grids. For a density grid g that is composed of $S_{1,j_1} \times S_{2,j_2} \cdots \times S_{d,j_d}, j_i = 1, \ldots, p_i$, we denote it as

$$g = (j_1, j_2, \cdots, j_d). \quad \text{--------------(3)}$$

A data record x = (x1, x2, · · ·, xd) can be mapped to a density grid g(x) as follows:

$$g(x) = (j_1, j_2, \cdots, j_d)$$ where xi 2 Si,ji .

For each data record x, we assign it a density coefficient which decreases with as x ages. In fact, if x arrives at time tc, we define its time stamp T(x) = tc, and its density coefficient D(x, t) at time t is

$$D(x,t) = \lambda^{t-T(x)} = \lambda^{t-t_c}, \quad \text{----------(4)}$$

where λϵ (0, 1) is a constant called the decay factor.

Definition (Grid Density) For a grid g, at a given

time t, let E(g, t) be the set of data records that are map to

g at or before time t, its density D(g, t) is defined as the sum

of the density coefficients of all data records that mapped

to g. Namely, the density of g at t is:

$$D(g,t) = \sum_{x \in E(g,t)} D(x,t).$$

1) *procedure initial clustering (grid list)*
2) *update the density of all grids in grid list;*
3) *assign each dense grid to a distinct cluster;*
4) *label all other grids as NO CLASS;*
5) *repeat*
6) *for each cluster c*
7) *for each outside grid g of c*
8) *for each neighboring grid h of g*
9) *if (h belongs to cluster c')*
10) *if (|c| > |c'|) label all grids in c' as in c;*
11) *else label all grids in c as in c';*
12) *else if (h is transitional) label h as in c;*
13) *until no change in the cluster labels can be made*
14) *end procedure*

Figure 3: The procedure for initial clustering.

1) *procedure adjust clustering (grid list)*
2) *update the density of all grids in grid list;*
3) *for each grid g whose attribute (dense/sparse/transitional) is changed since last call to adjust clustering()*
4) *if (g is a sparse grid)*
5) *delete g from its cluster c, label g as NO CLASS;*
6) *if (c becomes unconnected) split c into two clusters;*
7) *else if (g is a dense grid)*
8) *among all neighboring grids of g, find out the grid h whose cluster ch has the largest size;*
9) *if (h is a dense grid)*
10) *if (g is labeled as NO CLASS) label g as in ch;*
11) *else if (g is in cluster c and |c| > |ch|)*
12) *label all grids in ch as in c;*
13) *else if (g is in cluster c and |c| <= |ch|)*
14) *label all grids in c as in ch;*
15) *else if (h is a transitional grid)*
16) *if ((g is NO CLASS) and (h is an outside grid if g is added to ch)) label g as in ch;*
17) *else if (g is in cluster c and |c| >=|ch|)*
18) *move h from cluster ch to c;*
19) *else if (g is a transitional grid)*
20) *among neighboring clusters of g, find the largest one c' satisfying that g is an outside grid if added to it;*
21) *label g as in c';*
22) *end for*
23) *end procedure*

Figure 4: The procedure for dynamically adjusting clusters.

The calculated values of z1, z2, z3 components will be applied as an input for both the clustering algorithms to form the clusters based on their risk level.

*3) Patient's Health status*

Using clustering algorithm we form the clusters for attributes stated above. And then for patient's current input we predict patient's health status i.e. patient is fit or unfit.

IV. EXPERIMENTAL RESULT

The above described algorithms used for formation of clusters on medical database. The data will be collected from the Switzerland data set. The data set contains the 107 instances and the 14 attributes. The attributes are age, sex, Blood Pressure, Cholesterol, Chest Pain and etc. The performance of these algorithms will be computed by using correctly predicted instance. [1]

Performance Accuracy= correctly predicted Instance/
Total Number of Instance

TABLE I. PERFORMANCE OF CLUSTERING ALGORITHM

| Cluster Category | Cluster Algorithm | Measures | | |
|---|---|---|---|---|
| | | Correctly Classified Instance | In correctly Classified Instance | Prediction Accuracy |
| Clusters | Simple K-means | 89 | 18 | 83.18 |
| | D-stream | 94 | 13 | 87.85 |

From above table we observed that the performance of density based algorithm is better than simple K-means. Accuracy of D-stream algorithm is more than K-means.

## V. FUTURE SCOPE AND CONCLUSION

K-means is unable to handle arbitrary cluster formation because prediction of the number of classes to be formed is not fixed. The D-stream algorithm has superior quality and efficiency, can find clusters of arbitrary shapes, and can accurately recognize the evolving behaviors of real-time data streams. Therefore, D-stream will perform better in biomedical applications. This system can be further developed for real time analysis of biomedical data to predict patient's current health status.

The proposed system can be used for monitoring elderly people, Intensive Care Unit (ICU) Patient. Also the system gives the health status of patient, it can be used be used by clinicians to keep the records of patients.

The proposed system is adaptive since it can handles more than one physiological signal. The proposed system uses historical biomedical data which is very useful for prediction of current health status of a patient by using clustering algorithms like K-means, D-stream, etc. Prediction of health status is very sensitive job, D-stream will perform better here, as it supports arbitrary cluster formation which is not supported by K-means. Also D-stream is particularly suitable for users with little domain knowledge on the application data that means it won't require the K-values. Hence D-stream is parameter free and proves to give more accurate results than K-means when used for cluster formation of historical biomedical data.